\documentclass{eptcs}
\providecommand{\event}{Refine 2018} % Name of the event you are submitting to
\usepackage{underscore}     % Only needed if you use pdflatex.

\usepackage{amsmath}
\usepackage[dvipdfmx]{graphicx}
\usepackage{bmpsize}
\usepackage{oz-refine18}
\usepackage{xcolor}

\newcommand{\wmmRef}{weak-memory trace refinement\ }
\newcommand{\WMMRef}{Weak-memory trace refinement\ }

\newcommand{\PEv}{ProgEvents}
\newcommand{\ObjEv}{ObjEvents}

\newcommand{\sem}[1]{\lbag #1 \rbag}
\newcommand{\semM}[1]{\lbag #1 \rbag_M}

\newcommand{\order}[1]{\ <_{#1}\,}
\newcommand{\Porder}{\ <_{P_M}\,}

\newcommand{\res}[2]{#1\,_{|#2}\,}  
\newcommand{\obs}[1]{#1\,_{|global}\,}

\newcommand{\refsto}{\sqsubseteq}

\newcounter{axiom}

\renewenvironment{proof}{\par\noindent\textbf{Proof}\\}{\vspace*{-2ex}\\\hspace*{\textwidth}\hspace*{-6ex}$\Box$}

\newtheorem{lemma}{Lemma}
\newtheorem{definition}{Definition}

\title{Correctness of Concurrent Objects under Weak Memory Models}
 \author{Graeme Smith \and Kirsten Winter \and Robert J.\ Colvin
 \institute{
School of Information Technology and Electrical Engineering,\\
The University of Queensland, Australia
}
\email{smith@itee.uq.edu.au \quad kirsten@itee.uq.edu.au \quad  r.colvin@uq.edu.au}
}
\def\titlerunning{Correctness of Concurrent Objects under Weak Memory Models}
\def\authorrunning{G. Smith, K. Winter \& R. J. Colvin}

\begin{document}

\maketitle

\begin{abstract}

  In this paper we develop a theory for correctness of concurrent
  objects under weak memory models.  Central to our definitions is the
  concept of \emph{observations} which determine when effects of
  operations become visible, and hence determine the semantics of
  objects, under a given memory model.  The resulting notion of
  correctness, called \emph{object refinement}, is generic as it is
  parameterised by the memory model under consideration. Our theory
  enforces the minimal constraints on the placing of observations and
  on the semantics of objects that underlie object refinement.  Object
  refinement is suitable as a reference for correctness when proving
  new proof methods for objects under weak memory models to be sound
  and complete.

\end{abstract}

\section{Introduction}\label{sec:intro}

Linearizability \cite{HeWi90} is widely accepted as the standard
correctness criterion for concurrent objects, i.e., objects designed
to be accessed simultaneously by multiple threads \cite{Moi07}. 
 Recent work \cite{bur12,got12,ifm,DBLP:conf/hvc/TravkinMW13,DohertyDerrick2016,DongolVMCAI2018}
 has begun examining the applicability of linearizability in the
 context of weak memory models of modern multicore architectures
 \cite{Owens2009, Sewell:2010:XRU:1785414.1785443, Alglave:2014:HCM:2633904.2627752,armv8,ColvinFM2018}. These
 memory models improve hardware efficiency by limiting accesses to
 global memory. Individual threads operate on local copies of global
 variables, updates to the global memory being made by the hardware
 and largely out of the programmer's control\footnote{A programmer can add fences (or memory barriers) to code to force any pending updates to be written to memory. However, if used indiscriminately, fences cause the code to be less efficient.}. This can cause threads
 executing on different cores to get out of sync with respect to the
 values of global variables.

%-- Behaviour of WMM

 For example, on the TSO (Total Store Order) architecture
 \cite{Owens2009,Sewell:2010:XRU:1785414.1785443} a thread updating a global
 variable $x$ stores the new value in a per-core FIFO buffer. Threads
 executing on that core will then read $x$ from the buffer, rather
 than the global memory, until the new value is flushed from the
 buffer to global memory by the hardware. In the meantime, threads on
 other cores read the value of $x$ from the global memory or from
 their own core's buffer when it has a value for $x$. 
 %This is often described as a reordering of load instructions (reads) with write instructions.  

The situation is even more complex for weaker architectures, like ARM
and Power \cite{Alglave:2014:HCM:2633904.2627752,armv8,ColvinFM2018}. 
Writes to variables are, like on TSO, local to the core on which they occur and
are made available to other cores by the hardware. 
%or the use of fence instructions in the program. 
However, they are not necessarily made
available in FIFO order. A write to variable $x$ may be made available
globally after a write to variable $y$ although the write to $x$
occurs before the write to $y$ in program order.
%; the write to $x$ and the write to variable $y$ seem to occur out of order.

\begin{figure}[t]
   \centering
     \includegraphics[viewport=20 656 368 781]{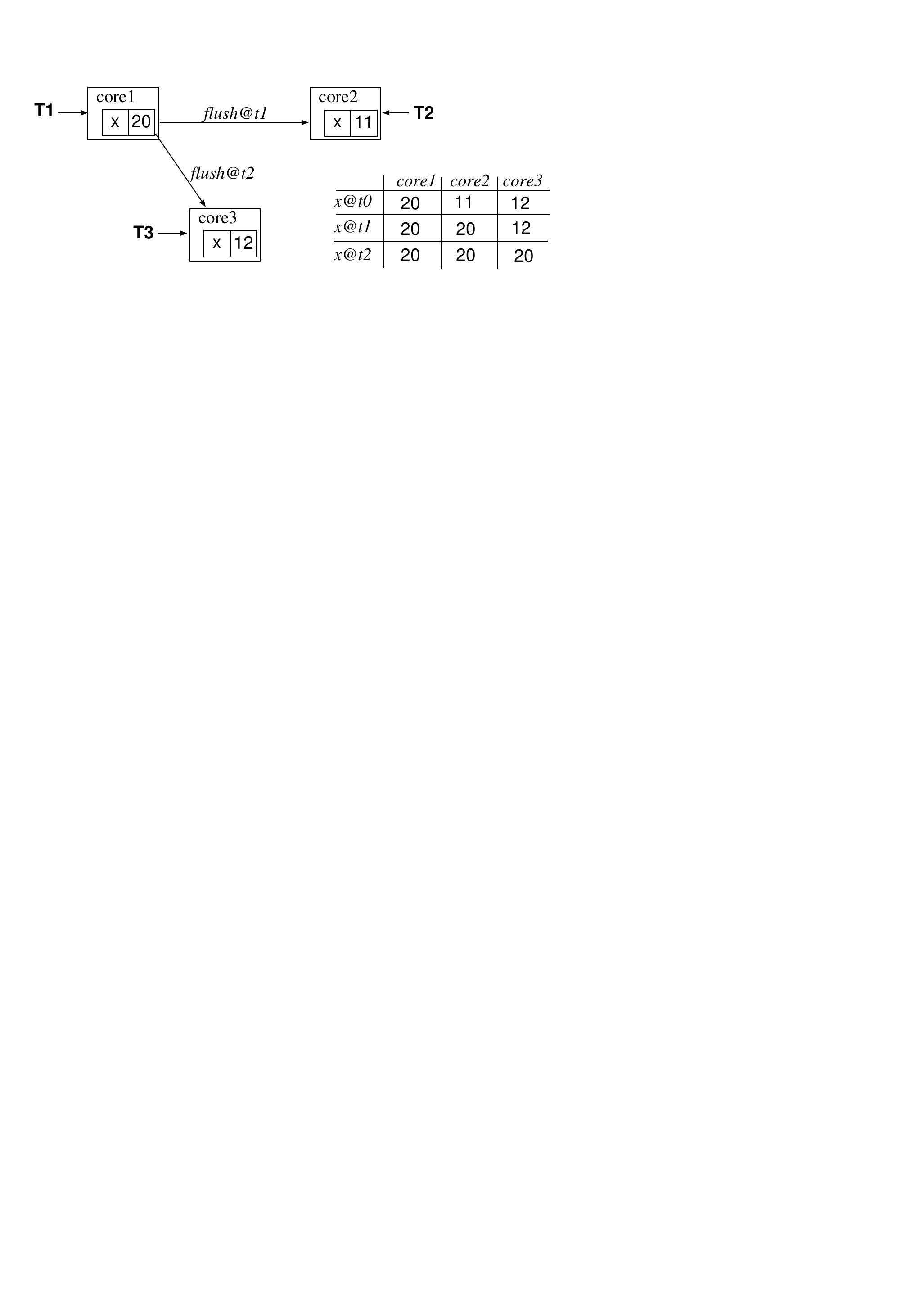}
   \caption{Scenario for non-multi-copy atomicity in ARM and Power architectures}
   \label{fig:cores}
 \end{figure}

 Additionally, values of global variables are flushed
 %independently (and 
 at different times to different cores (this is
 referred to as \emph{non-multi-copy atomicity}). 
 This is illustrated in Figure~\ref{fig:cores} where three different cores are depicted
with one thread running on each
and referring to a global variable $x$. Assume 
at time $t\,0$ thread T1 performs a write to $x$, which then gets flushed
to core2 at time $t1$, and later at time $t2$ to core3. The evaluation of $x$
per core at these time instances is depicted in the table, which shows that for this scenario
the value is consistent only at time $t2$ (for scenarios other than the depicted one,
there might not be a consistent state at all).
% A write to variable
% $x$ may be made available globally after a write to variable $y$ even though
% the write to $x$ occurs before the write to $y$ in program order.
% This can be perceived as a reordering of write instructions within one thread if 
% they address different variables. 

To evaluate proposed notions of linearizability for weak
memory models we need to argue their soundness and completeness.
%(and even their compositionality). 
That is, we need to prove that an object
implementation linearizes to its specification if and only if it is
\emph{correct} with respect to its specification. However, this
requires a notion of correctness as a reference point.  What does it
mean for an implementation to be correct when executed on a weak memory
model?

Traditionally, trace refinement provides this notion of correctness
for programs: an implementation is correct with respect to its
specification if and only if each observable behaviour of the
implementation can also be observed from the specification \cite{bac94a,back,abadi}. 
In the context of concurrent objects that are called by a \emph{client program}, 
an object implementation is deemed correct if and only if the client program cannot differentiate between 
the object implementation and its abstract specification, as the observable behaviour is the same.

To capture what is observed by client programs under \emph{sequentially consistent} (SC) architectures (i.e., those without a weak memory model), 
the notions of \emph{observational refinement} \cite{Filipovic-LinvsRef2010} and 
\emph{contextual refinement} \cite{Dongol2016} have been
introduced. However, both of these definitions do not provide a notion
of correctness under weak memory models, as they do not take into account
that an event occurring on one core might be observable later on another. Instead
events become observable immediately after their occurrence. 
For architectures with non-multi-copy atomicity, in which flushes occur out of write order,
the definitions do not provide us with the right semantics of programs.

%-- Our approach

In this paper, we propose a general definition of correctness for
concurrent objects running on {\em any\/} existing memory model. The
aim is to provide a reference point for proving new notions of
object correctness 
%linearizability
 for weak memory models sound and complete.  The
result is a notion of object refinement
%version of trace refinement, called \emph{\wmmRef}, 
which is parameterised by the memory model it refers to, and  is therefore
generic and can be instantiated for any weak memory model
behaviour.  Key to this result is a definition of the semantics of an
object operating in the context of a calling client program under a weak
memory model.  In particular the semantics of an object's
specification in the context of a client program under a weak memory
model needs be defined in such a way that its behaviour maintains the
intention of the specification, namely that operations are atomic.
% despite the effects of the weak memory model.

 The paper is structured as follows.
In Section~\ref{sec:prog} we introduce the basic concepts of our theory
including that of \emph{observations} which is key to our definitions. 
Based on these concepts, Section~\ref{sec:progsemantics} formalises the semantics of 
programs under a given memory model.  
The semantics of a concurrent object in the context of a client, under
a memory model, is elaborated in Section~\ref{sec:objectsemantics},
distinguishing cases for the specification, in which operations are
atomic, and the implementation which includes non-atomic, and possibly
non-terminating, operations.
Section~\ref{sec:wmmref} ties these basics into the notion of \emph{\wmmRef}$\!\!$ which defines
refinement under memory model $M$ for a client program using an object and its specification, respectively.
The definition is parameterised with a given memory model for which we assume
a semantics is given. 
Using \wmmRef we can then define our notion of \emph{object refinement} under memory model $M$
which delivers the notion of correctness for objects.
Section~\ref{sec:ex}, illustrates how our
 definition can be used to prove correctness of a case study.
The paper concludes with a discussion of related work in Section~\ref{sec:relwork}.

\section{Client programs and observation events}
\label{sec:prog}

To investigate the behaviour of concurrent objects under weak memory models, 
and relate their implementation to their specification using refinement, 
we need to consider the calling context.
Programs calling the operations of a concurrent object are referred to as \emph{client programs},
or clients for short. A client program $P$ is concurrent, spawning multiple threads $T_i$
on multiple cores, and is affected by the memory model of the architecture it is running on. 
For some finite $n$, we have\footnote{For simplicity, we do not consider dynamically spawned threads.}
\[ P \sdef T_1 \parallel T_2 \parallel \ldots \parallel T_n .\]

Following other work on concurrent objects \cite{HeWi90, Filipovic-LinvsRef2010, Dongol2016}, the behaviour of a program is
described in terms of \emph{events} that occur. We allow events to be \emph{program steps}, \emph{operation events} or, as introduced in Section~\ref{sec:observations}, \emph{observation events}.

\emph{Program steps} are steps, other than calling an object operation, performed by the client.
These are assignments, conditional branch instructions (e.g., {\sf if} 
or {\sf while} statements), other control instructions like various forms of fences,
atomic read-write-modify instructions which atomically perform these three steps 
(e.g., the compare-and-swap construct CAS),
and higher-level instructions which can, in many cases, be defined in
terms of assignments and/or conditional branches. For example, a statement
{\sf await(z=1)} could be defined as {\sf while(z $\neq$ 1)\,$\{\}$}.

\emph{Operation events} abstract the effects of an operation call by a program. They include the \emph{invocation} of the operation (i.e., when it is called) and
the operation's \emph{response} (i.e., when it returns). The operation events carry the operation's input and output values
as parameters and thus reflect the operation's externally visible behaviour. The internal 
behaviour of an object is elided.  

All program steps and operations are deterministic; non-determinism in
a program results from the interleaving of events on different threads
varying between executions.

%Programs might contain \emph{synchronisation points} 
%to reduce the possible interleavings between events of different thread,
%as the threads are forced to ``synchronise''.
%For example, an {\sf await} statement in one thread
%synchronises that thread with a different thread that contains the
%event that is waited for.  Synchronisation points are the only means
%to enforce an order between events of different threads.  Note, that
%we do not consider fence instructions as synchronisation points
%since fences only synchronise flushes but not other program steps or object events.

\subsection{Observation events}
\label{sec:observations}

Central to our definitions is
the notion of an additional type of event called an {\em observation event\/}. 
Such an event denotes the point in an execution where 
a program step or the response of an operation 
can be deemed to have been observed by \emph{all} threads $T_i$ of client $P$. 
Such observed events are either 
\begin{itemize}
\item program steps that write to global program variables, or
\item responses of operations that write to object variables 
 that are shared by threads accessing the object.
\end{itemize}
Note that generally operation responses as such are not observable, as 
the returned value (if any) is written to a thread-local variable only. 
A subsequent program step is required to write it to a global program 
variable \cite{Filipovic-LinvsRef2010,Smith2017}.
%Note further that only the program steps are observed by the client program $P$. 
Also operation internal steps and their effects are only observable on the
object level but not on the program level (due to object encapsulation).  
Program refinement only takes into account observable program steps, 
and hence it is sufficient to consider only those.  
However, the observation of operations become essential when defining the semantics of
objects in a context of a client running under weak memory model, 
as we will see in Section~\ref{sec:objectsemantics}.

Introducing observation events allows us to decouple the occurrence of an event
and its observation which might not fall together under weak memory models. 
The semantics specific to the memory model under consideration
determines the possible placings of observation events. 
For example, the operational semantics for TSO given in \cite{Owens2009}, 
or for ARM and Power (e.g., \cite{ColvinFM2018}),
can produce all possible executions from which we can deduce the points of observations
according to the following rules:
Generally, an observation will occur when \emph{all} threads can either

\begin{enumerate}
\item[(a)] access a new value of a global program variable written by a program step, 
\item[(b)] access the values of \emph{all} shared object variables written by an operation, 
           \emph{after} the response of the operation has occurred, or
\item[(c)] %this case has changed: due to the soundness proof in which reordered concrete traces need to 
           %match an abstract trace in which the operation is atomic (hence obs follows res directly)
           when the response of a \emph{covert} operation, whose steps or outcomes are not directly visible, 
           has occurred.
\end{enumerate}
Covert operations (referred to in case (c))
are those which do not write to any shared object variables and either have no return value or
have a return value which is not written to a global program variable
(\cite{Filipovic-LinvsRef2010} does not consider these operations), and as such they are not 
directly observable.
However, from the observation of an event that must have occurred later than the response of the covert operation, we can deduce
that the covert operation must have occurred, and it
is thus indirectly observable. To ensure those two observations are ordered correspondingly,
the observation of a covert operation is placed directly after its response.
Examples of cases (a) and (b) are simply the flushes of global program and shared object variables
(of which we consider only the flush to the last core when non-multi-copy atomicity is present). 
If an operation writes to multiple shared object variables, the observation of the last write is considered.
Note that observations of writes in an operation (as in case (b)) 
or covert operation responses (as in case (c)) can only occur after the response of the operation.

\subsection{Examples}
\label{sec:prog-ex}

The effects of concurrency (interleaving the behaviour of threads) as
well as the effects of weak memory models leads to a partial ordering
of the events of a program to which all possible executions of that
program must adhere.

 \begin{figure}[b]
   \vspace{-8ex}
 \[
   \qquad \qquad\qquad\qquad\parbox[t]{3.2cm}{~~{\sf T1} \vspace*{2mm}\\ 
      {\sf x := object.A();\\ z := 1;\\ object.B();}} 
%     \parbox[h]{1cm}{\vspace*{2cm}{$\zBig\parallel$}} 
     \parbox[h]{1cm}{\vspace*{2.2cm}{\resizebox{!}{0.5cm}{$\parallel$}}} 
      \parbox[t]{3cm}{~~{\sf T2}\vspace*{2mm}\\ {\sf await(z=1);\\ object.C();}}
 \]
 \caption{Client program}
  \label{fig:client}
 \end{figure}

Consider the example of a client program given in Figure~\ref{fig:client}
with two global program variables {\sf x} and {\sf z}, running on two threads {\sf T1} and {\sf T2}. 
The threads call three operations of the same object.
%Assume operation {\sf A} writes to a shared object variable and operations {\sf B} and {\sf C} read this object variable. 
%
 Note that we assume that the result of any operation call is implicitly 
 stored in a local register (e.g., {\sf r}$_{\sf A}$ for the operation call
 {\sf object.A}), and ${\sf inv_A}$ denotes the invocation of {\sf object.A}, etc.

 On an SC architecture (i.e., one
 without a weak memory model), writes to global variables occur
 instantaneously. Hence, the observation event for a program step
 occurs immediately after the program step, and that of an operation
 immediately after its response. 
 The partial order of the events of the program of
 Figure~\ref{fig:client} on SC is shown below.

\begin{center}
 \scalebox{.6}{\includegraphics[viewport=4 703 589 832]{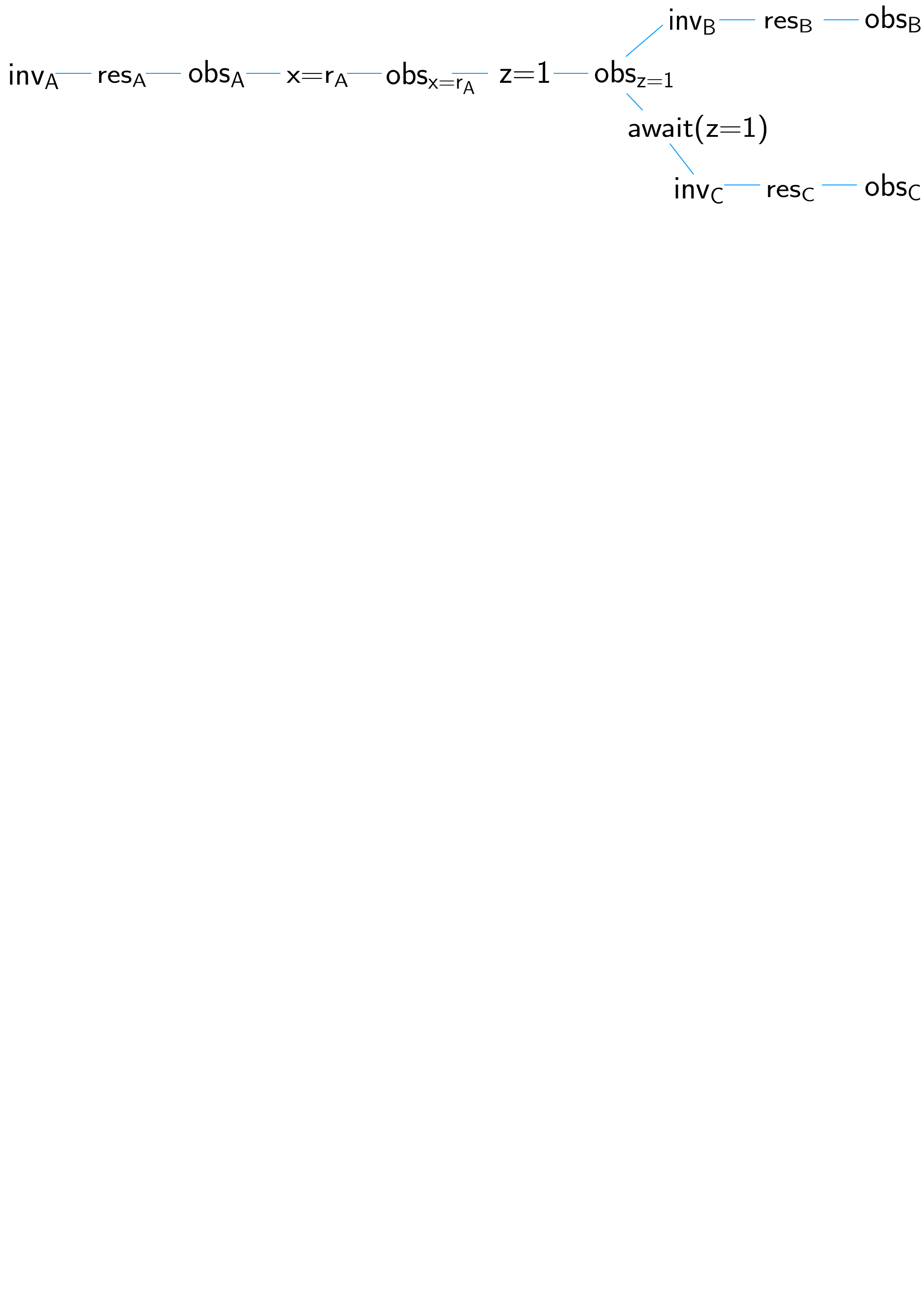}}
\end{center}

On TSO, writes to global variables and shared object variables 
become available to threads on other cores when they are flushed. 
Furthermore, flushes occur in the same order as the writes occurred.
Hence, in the partial order of the events on TSO (depicted below),
${\sf obs_{A}}$ occurs before ${\sf obs_{x=rA}}$ which occurs before
${\sf obs_{z=1}}$ which occurs before ${\sf obs_{B}}$.
In contrast, the observations ${\sf obs_{B}}$ and ${\sf obs_{C}}$ are un-ordered 
as they occur on different threads.

\begin{center}
 \scalebox{.6}{\includegraphics[viewport=4 656 484 797]{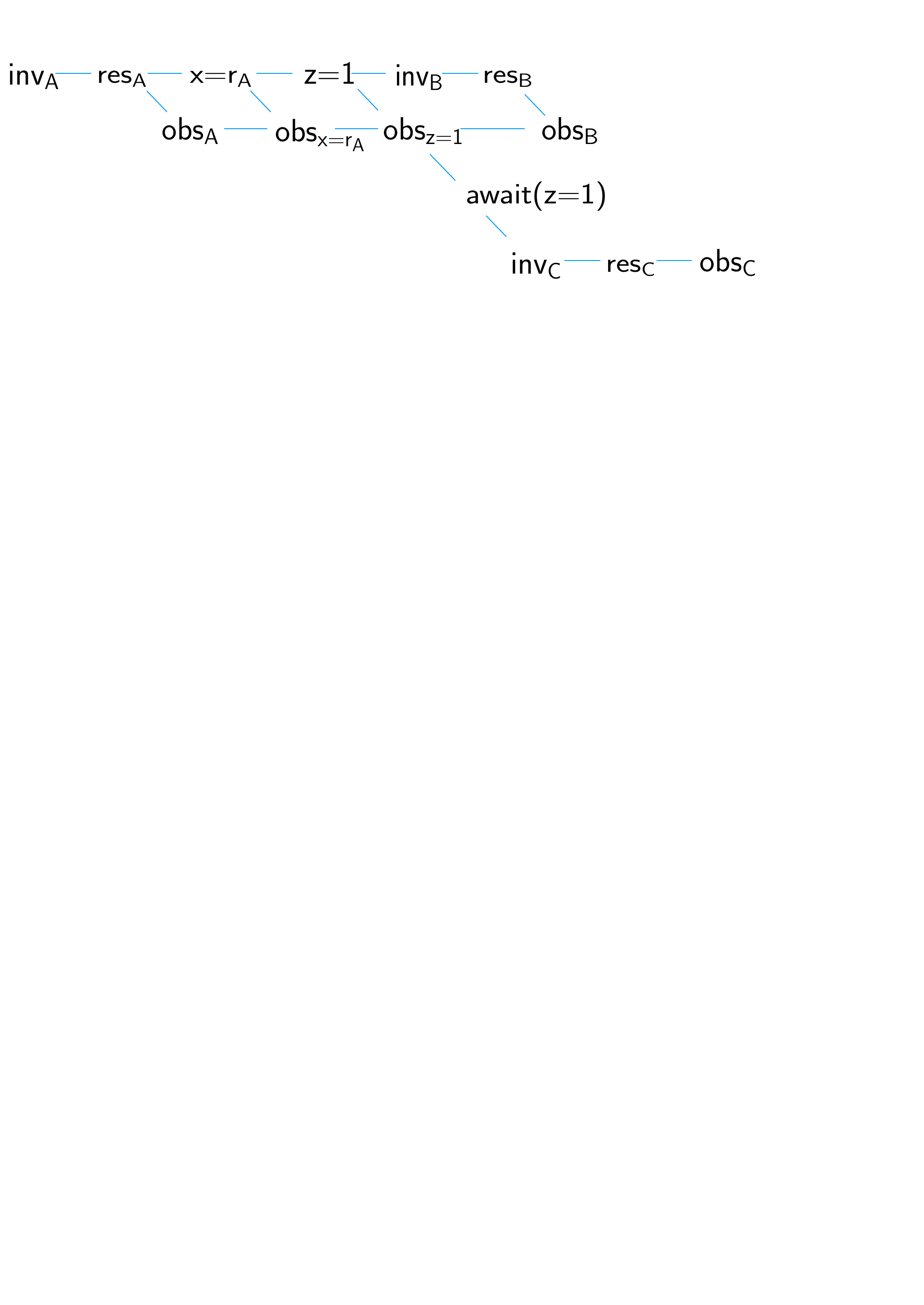}}
\end{center}

On ARM and Power, flushing of variables does not occur in a FIFO manner,
hence the observations are un-ordered in the partial order of the
program depicted below. The partial order is split into four
sub-graphs which are unrelated if we assume that there are no data dependencies 
on shared object variables between the operations. Consequently the 
object events can occur out of order for as long as the object events per operation
adhere to the wellformedness condition of traces.\footnote{This may lead to 
operations on a single thread overlapping.}
The assignment to {\sf x}  and  the {\sf await} statement maintain their order with respect to their immediately preceding events,
the former due to a data dependency which prohibits reordering, the latter due to 
the synchronisation between {\sf T1} and {\sf T2}. 

\begin{center}
 \scalebox{.6}{\includegraphics[viewport=4 612 238 797]{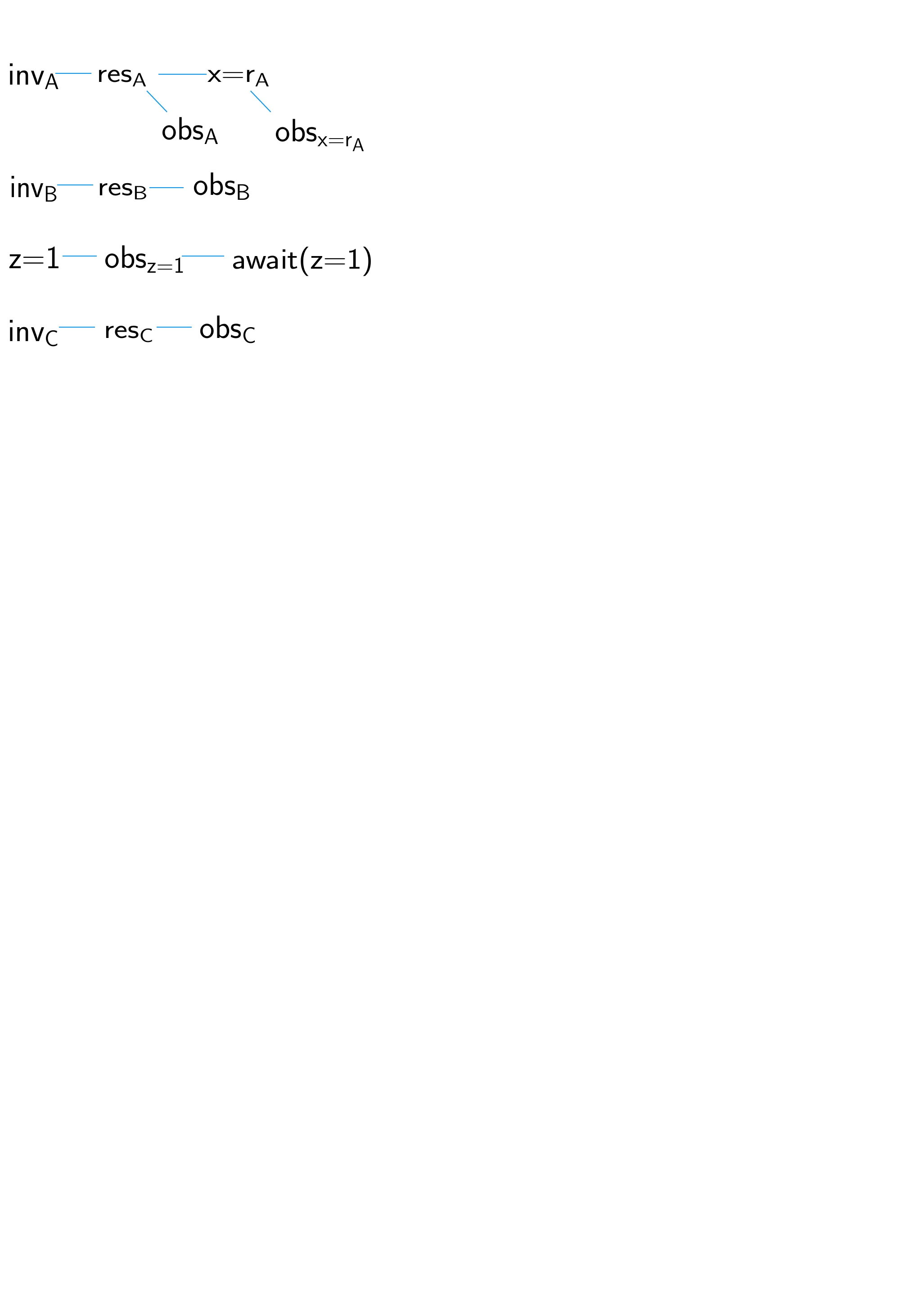}}
\end{center}

\section{Semantics of programs relative to the memory model}\label{sec:progsemantics}

The semantics of a program $P$ under  memory model $M$ is defined in terms of
the set of {\em events\/} that can occur, and the {\em partial order\/} over those events. 
%
%Note that 
%the partial order, and in particular the placement of observation events, is determined 
%by the memory model as outlined in Section~\ref{sec:prog}.
%
The partial order of events is partly enforced by $P$ through the
program text; this is referred to as the \emph{program
  order}. Additionally, in particular the placement of observation events, is determined by 
  %it reflects 
  the semantics of the memory
model as outlined in Section~\ref{sec:prog}. For example, under TSO the order of writes to variables is
determined by the program order, and the order of observations of such
writes is the same as the order of the writes (due to the FIFO nature
of the store buffer).
We formalise the semantics of programs as follows. 

\subsection{Events}

Let $T$ be the set of all thread ids, and $Call$ the set of all
operation calls. An operation is then defined as a call by a
particular thread.

\[Op\sdef T\cross Call\]

Let $PS$ denote the set of all program step events, and $Val$ the set
of all values (of input and output parameters) including a special
element $\bot$ meaning `no value'. The set of all events is defined as
follows where each invocation is associated with an input, and each
response and observation with an output.

\[Event  \sdef  step(T,PS) | obs(T,PS) | inv(Op,Val) | res(Op,Val) | obs(Op, Val) \]
In the remainder of this paper, we refer to $step(T,PS)$ and $obs(T,PS)$ as program events, $\PEv$, 
and to $inv(Op,Val)$, $res(Op,Val)$, and $obs(Op, Val)$ as object events, $\ObjEv$.

A program $P$ has a set of events, $events(P)$, such that for each
invocation event in the events of $P$, $events(P)$ also contains the corresponding response and
observation events. That is, each called operation can respond and be observed.

\[\all op:Op; in:Val\dot inv(op,in)\mem events(P) \implies\\
\t1 \exi out:Val\dot \{res(op,out),obs(op,out)\}\subseteq events(P) & (\refstepcounter{axiom}\arabic{axiom})
\label{events}
\]

\subsection{Traces}

The semantics of a program $P$ is described as a set of {\em finite\/} sequences of
events, referred to as {\em traces\/}.\footnote{Since we are interested in
defining a notion of correctness that readily relates to linearizability, 
and hence only safety properties \cite{GotsmanYang2011,Smith2017}, we do not
consider infinite sequences of events in our semantics.} 
For each trace
$t$, each event is unique (similar events, e.g., calls to the same
operation, may be annotated by their relative position in the trace),
and an invocation of an operation always occurs before the associated
response, which in turn occurs before the associated
observation. Similarly, a program step always occurs before its
observation, if any. In the following $s_i$ denotes the $i$th element of a sequence $s$, and $\#s$ its length.

\[Trace \sdef \M \{t:\seq Event| (\all i,j \leq \# t\dot i\neq j \implies t_i\neq t_j) \land\\
\t1 \all c:Op\cross Val\dot \M(\all j\dot t_j=res(c) \implies \exi i < j \dot t_i=inv(c)) \land\\
(\all j\dot t_j=obs(c) \implies \exi i < j \dot t_i=res(c))\land\O\\
\t1 \all s:T; p:PS\dot(\all j\dot t_j=obs(s,p) \implies \exi i < j \dot t_i=step(s,p))\}\O\]

The events of a trace and the order on these events are defined as
follows. Note that the order $\order{t}$ is a total order over the
events in $t$, as a trace describes exactly one execution.
\begin{align*}
events(t) &\sdef \{a:Event| \exi i\dot t_i=a\}\\
\order{t} &\sdef \{(a,b):Event\cross Event | \exi i,j\dot i < j \land t_i=a \land t_j=b\}
\end{align*}

\subsection{Partial order}
%----  Porder   --------------------------------------------

The semantics of program $P$ on memory model $M$ is then defined as
the set of traces using only events from $P$ and whose orders adhere to
the constraints prescribed by the partial order
$\!\!\Porder\!\!$. 

\[\semM{P} \sdef \{t:Trace|events(t) \subseteq events(P) \land \Porder \Subset \order{t}\}\]
where $\Porder \Subset \order{t}$ specifies whether an order is \emph{allowed} by $P$ on $M$,
formally defined as  

\[\Porder \Subset \order{t} \sdef \all (a,b):\Porder\dot b\mem events(t) \implies (a,b) \mem \order{t}\]
That is, for any event $b$ that occurs in trace $t$, if this event is
enforced to come after another event $a$ by $\!\!\Porder\!\!$, then
event $a$ must have also occurred in $t$ before event $b$. Note that
it is not suitable to use the simple subset relation here, i.e.,
$\Porder \subseteq \order{t}$, since trace $t$ will not, in general,
include all events $b$ that are restricted by $\!\!\Porder\!$.

Since we are aiming at the most general description of program semantics
under any memory model, we do not explicitly prescribe
$\!\!\Porder\!\!$ but based on the understanding of concurrent
programs, concurrent objects and their interplay (as laid out in Section~\ref{sec:prog}) we can formulate certain characteristics of
$\!\!\Porder\!\!$ that are shared by all memory models, and provide a set of axioms 
in the remainder of this section.

%----  Assumptions on P and O  --------------------------------------------

In the context of this work, the order of operation events in $\!\!\Porder\!\!$
are of particular interest together with the order between operation
and program events.  We refer to the notion of \emph{synchronisation}
(between threads) as an event on one thread that affects the occurrence of events
on another thread, and hence affects the overall order of events. 
Synchronisation  requires a writing and a reading access to a shared variable 
(e.g., through an {\sf await} statement or a conditional).
We make the following assumptions on objects and clients:
\begin{itemize}
\item The state space of an object is encapsulated and hence the client does not share any 
  variables directly with the object; the communication occurs only through input and output values of the objects
  operations.
\item Consequently, synchronisation between two threads can only occur between
  program events (referred to as \emph{program synchronisation})
  or between operation events  (referred to as \emph{object synchronisation}). The client cannot directly
  synchronise with a step in an operation that occurs between its invocation and
  response but only with the invocation and response events themselves which serve as an interface
  between client and object. 
\item A client cannot enforce a flush within an operation; 
  the operations implementation is outside the client's control.
\end{itemize}

\begin{figure}[t]
 \centering
  \scalebox{0.9}{
  \includegraphics[viewport=7 760 388 825]{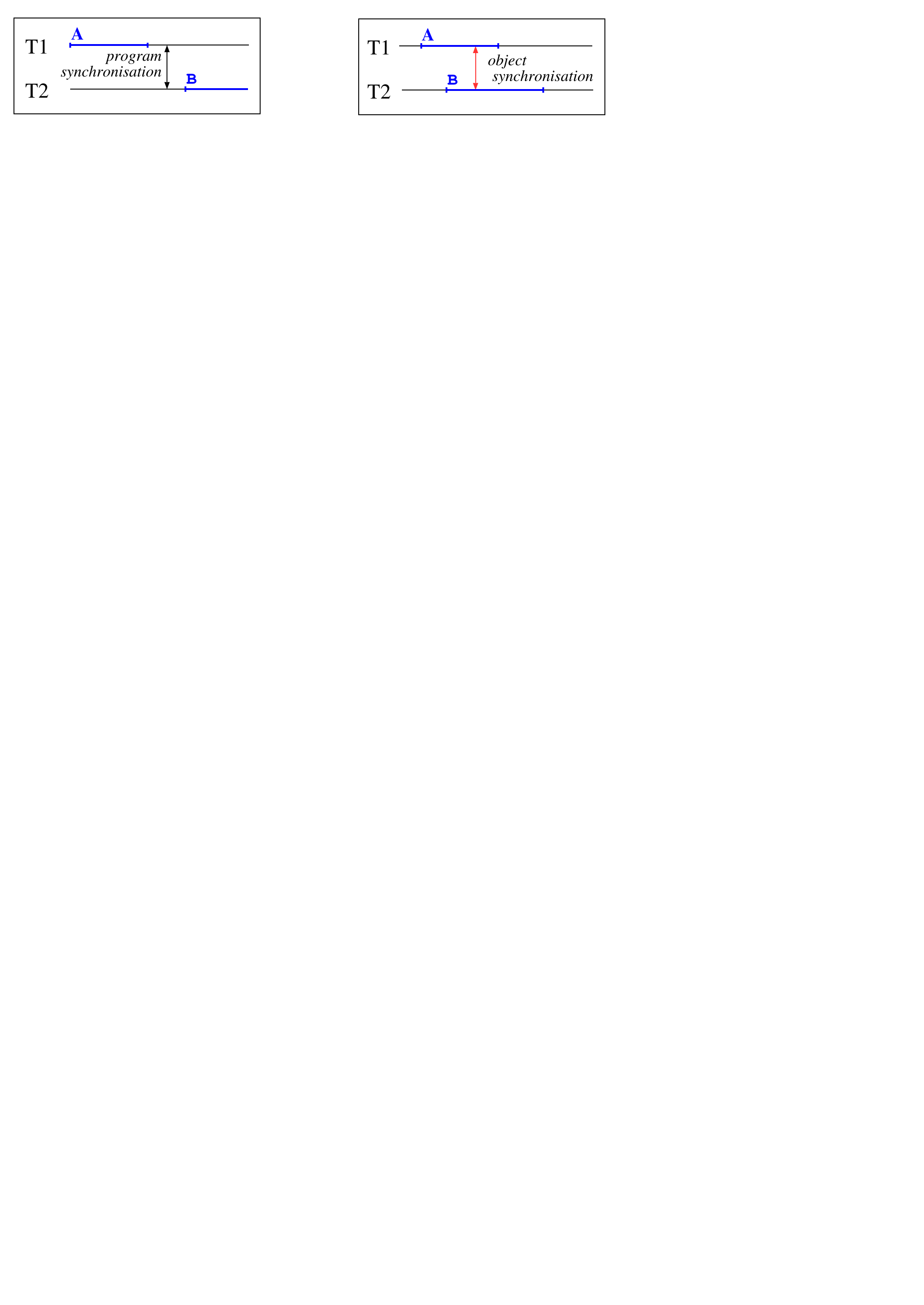}}
 \caption{Scenarios of synchronisations between threads}
 \label{fig:sync}
 \end{figure}

Figure~\ref{fig:sync} gives an intuition for the possible synchronisations by means of two scenarios
  of threaded behaviours of a client calling two operations {\sf A} and {\sf B}. The scenario 
on the left illustrates program synchronisation between the threads {\sf T1} and {\sf T2},
in which the program synchronises on two program steps which follow operation {\sf A} and precede
operation {\sf B}, respectively. The second scenario on the right illustrates operation
{\sf A} and {\sf B} overlapping in which case the synchronisation between them must be an object synchronisation
which is outside the control of $P$. % under $M$. 

From these assumptions we can deduce that 
$\!\!\Porder\!\!$ can only enforce the invocation of an operation to occur
\emph{before} another event if and only if also the response is enforced to occur before the event.
(The implication from right to left holds due to the wellformedness condition on traces.)

\[\all e: Event; c : Op\cross Val\dot 
    (inv(c), e)\mem \Porder 
    \iff
    (res(c), e)\mem \Porder & (\refstepcounter{axiom}\label{enf-inv-before}\arabic{axiom})
\]

Similarly, $\!\!\Porder\!\!$ cannot enforce the response of an operation
to come \emph{after} another event unless it also enforces the invocation to come after the event.
(As above, the implication from right to left holds due to wellformedness.)

\[\all e: Event; c : Op\cross Val\dot 
    (e, res(c))\mem \Porder 
    \iff
    (e,inv(c))\mem \Porder & (\refstepcounter{axiom}\label{enf-res-after}\arabic{axiom})
\]

A similar result holds for the order of observation events with respect to
program events: Due to an object's implementation, the oberservation of an
operation might occur directly after its response event (e.g., if the operation
is implemented with fence instructions which prevent a delay in its
oberservation, or if the operation is a covert operation), which is outside the control of $P$.
% or $M$.  
Hence
$\!\!\Porder\!\!$ cannot enforce the observation of an operation to come
\emph{after} another \emph{program} event unless it also enforces the invocation
to come after the program event.
(Again, the implication from right to left holds due to the wellformedness condition on traces.)

\[\all e: \PEv; c : Op\cross Val\dot 
    (e, obs(c))\mem \Porder 
    \iff
    (e,inv(c))\mem \Porder & (\refstepcounter{axiom}\label{enf-obs-after-ps}\arabic{axiom})
\]

However in some memory models, the order in which two observations can occur can be constrained 
(e.g., in TSO the order of observations follows the order of observed events), but only
if the order of the observed events is also enforced.
This means that the order of the observed operations must be
enforced, and hence the response of one operation must occur before the 
invocation of the other. 

\[\all c,d:Op\cross Val\dot 
    (obs(c), obs(d))\mem \Porder 
    \implies 
    (res(c),inv(d))\mem \Porder 
\]

A similar result also holds if $\!\!\Porder\!\!$ enforces an operation observation to occur after
the invoke or response of another observation. This is only possible if the two
operations are ordered (and hence the response of the first must occur before
the invocation of the other). If the operations are not ordered by $\!\!\Porder\!$, they may occur in
either order (and hence the observations may occur in either order) 
or they may overlap. For overlapping operations (necessarily executing on two different threads),
$\!\!\Porder\!\!$ cannot enforce an order of object events since any object synchronisation is beyond the control
of $P$. Therefore we can generalise the above axiom as follows.

\[\all c,d:Op\cross Val\dot \all e : \{inv(c), res(c), obs(c)\}\dot
  (e, obs(d)) \in \Porder
  \implies
  (res(c), inv(d)) \in \Porder & (\refstepcounter{axiom}\label{enf-order-obs}\arabic{axiom})
\]

As a consequence of these observations we deduce that 
an order between object events of two operations can only be enforced by $\!\!\Porder\!\!$ if 
$\!\!\Porder\!\!$ also enforces that these two operations do not overlap. 

\begin{lemma}[Enforced ordering on object events]\label{porderLemma}
\[\all   c, d:Op\cross Val\dot \all e_1 : \{inv(c), res(c), obs(c)\}; 
        e_2 : \{inv(d), res(d), obs(d)\} \dot \\
    \qquad       ((e_1, e_2) \in \Porder  \land c \neq d)    ~\implies~ (res(c), inv(d)) \in \Porder
\]
\end{lemma}

The proof constitutes the simple application of Axioms  (\ref{enf-inv-before}) to (\ref{enf-order-obs})
 to all combinations of invocation, return and observation events of two operations.
%Figure~\ref{fig:sync} provides an intuition for this restriction on $\!\!\Porder\!$.

\section{Semantics of objects under weak memory models}
\label{sec:objectsemantics}

To define trace refinement between client programs using objects, we
need to constrain the behaviour of a program to a particular
object or collection of objects. If the program uses a collection of interacting objects,
we simply consider the collection of operations (and their events) provided by all objects 
in the collection. Below we consider a single object only.

The semantics of an object is given as a set of \emph{histories}, such that
each history is a trace with only object events,

\[History \sdef \{t:Trace|\res{t}{o}=t\}\]
where $\res{t}{o}$ denotes the trace $t$ restricted to only object events. 

\subsection{Object implementation under weak memory models}
\label{subsec:semC}

An object implementation $C$ has a set of object events, $events(C)$, 
and, on a particular memory model $M$, a prefix-closed set of histories 
made up of those events, $\semM{C}$. 
Observation events are not controlled by the object and hence can occur at any time 
after the associated response event that the memory model allows. 

For any object implementation $C$, $P[C]$ denotes the object $C$ operating in program $P$. 
It is only defined when
% $\{a:events(P)| \exi c:Op\cross Val\dot a\mem \{inv(c), res(c), obs(c)\}\}$ $\subseteq events(C)$, i.e., 
all object events of $P$ are events of $C$. 
The semantics of $C$ operating in $P$ on memory model $M$, $\semM{P[C]}$, is given as those traces 
of $P$ on $M$ whose object events correspond to a history of $C$ on $M$.

\[\semM{P[C]} \sdef \{t:\semM{P}|\exi h:\semM{C}\dot \res{t}{o}=h\}\]

\subsection{Object specification under weak memory models}
\label{subsec:semA}

An object specification $A$ similarly has a set of object events,
$events(A)$, and a prefix-closed set of histories, $\sem{A}$. Since
$A$ represents a typical specification found in a software library,
its set of histories is \emph{independent} of the memory model (hence there is no subscript).  
Any weak memory model behaviour is absent from its histories due to
%and since its operations are perceived as 
its operations being {\em atomic\/}, i.e., they occur 
without interference from other operations.

%To capture this in our semantics, in the histories of $A$ the positions of observation events is unrestricted,
%other than occurring after the corresponding response event. 
%
%%We denote with $\res{h}{\{inv,res\}}$ history $h$ restricted to only its invoke and response events.
%
%\[\all h:\sem{A}  \dot
%\exi c:Op\cross Val; i\dot h_i=res(c) \implies \\
%\t1 \all j > i \dot \exi h':\sem{A}\dot h \sim h' \land h'_j=obs(c)
%%\res{h}{\{inv,res\}} = \res{h'}{\{inv,res\}} \land h'_j=obs(c)
%& (\refstepcounter{props}\label{variant_obs}\arabic{props})
%\]
%where $h \sim h'$  denotes that $h$ and $h'$ are {\em thread equivalent\/}, i.e., when restricted to the events of any one thread they have the same sequence of invocations and responses.

\begin{figure}[t]
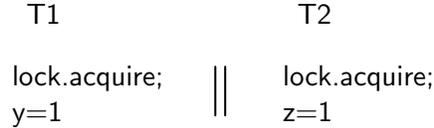

  \vspace{-8ex}
\[
 \qquad \qquad \qquad \quad \qquad\qquad\parbox[t]{2.6cm}{~~{\sf T1} \vspace*{2ex}\\ 
  {\sf lock.acquire;\\ y=1}} 
%   \parbox[h]{1cm}{\vspace*{1.6cm}$\zBig\parallel$} 
     \parbox[h]{1cm}{\vspace*{2cm}{\resizebox{!}{0.5cm}{$\parallel$}}} 
    \parbox[t]{3cm}{~~{\sf T2}\vspace*{2ex}\\ {\sf lock.acquire;\\ z=1}}
   % \qquad \qquad {\sf T1:}\qquad\parbox[t]{2.5cm}{\sf lock.acquire;\\ y=1}
   %  \parbox[h]{1cm}{$\zBig\parallel$} 
   %  {\sf T2:}\qquad\parbox[t]{2cm}{\sf lock.acquire;\\ z=1}
\]
\caption{Client program using lock}
 \label{fig:lock}
\end{figure}

To capture this in our semantics, the histories of $A$ are restricted
to those where only operations on one core are active at a time.
For example, suppose the specification of a lock object, {\sf lock},
has an operation {\sf acquire} which waits until the value of a
variable of the object, {\sf x}, is 1 and sets it to 0, i.e., {\sf
  acquire} is specified as {\sf await(x=1); x=0}. Assuming {\sf x} is
initially 1, in the program of Figure~\ref{fig:lock} the intention would
be that only one of {\sf y} or {\sf z} would be set to 1.

On SC, this intention is achieved when the invocation of the {\sf acquire}
which occurs second does not happen until after the response of the
{\sf acquire} which occurs first. On TSO and assuming T1 and T2 are
running on different cores, the intention is only achieved when the
invocation of {\sf acquire} which occurs second does not happen until
after the flush of {\sf x} from the {\sf acquire} which occurs
first. In both cases, the intention is met when the second occurrence
of {\sf acquire} is not invoked before the observation event of the
first occurrence. In general, for any object specification to behave
as intended, an operation invocation on one core does not occur
before the observation event of a previously invoked operation on any
other core. (The same does not need to hold for operations on the
same core, as values are read locally.) 

Let $core(a)$ denote the core on which an event $a$ occurs. We assume the core can be determined from the thread of the event. 

\[\all h:\sem{A}  \dot
%\t1 (\all i \leq \#h\dot h_i\mem events(A))\land\\
\all c:Op\cross Val; i \dot h_i=inv(c) \implies\\
\t2 \all op:Op;in:Val; j <i \dot 
%\t3 
h_j=inv(op,in)\land core(inv(op,in))\neq core(inv(c)) \implies\\
\t4 \exi out:Val; k< i\dot h_k=obs(op,out) & (\refstepcounter{axiom}\label{atomic}\arabic{axiom})\]

Additionally, since operations in a specification are intended to
always respond, for each history $h$ of $A$ with {\em pending\/}
invocations, i.e., invocations for which there is no response, the
history which extends $h$ with the missing events is also
in~$\sem{A}$. %'s histories.

\[\all h:\sem{A}; c:Op\cross Val \dot \exi i\dot h_i=inv(c) \land (\nexi j\dot h_j=res(c))\implies\\
\t1 \exi h\cat hr:\sem{A}; j\dot (h\cat hr)_j=res(c) & (\refstepcounter{axiom}\label{completions}\arabic{axiom})\]
This is not the case for object implementations which may have
operations which never return (e.g., due to an infinite loop).

Provided all object events of a program $P$ are events of a
specification $A$, $P[A]$ denotes the program $P$ operating with an
object whose behaviour satisfies~$A$.
The semantics of $P$ imposes restrictions from the 
memory model $M$ on to the traces and restricts the placement of observation events.
The semantics of $P[A]$ is given as follows.
\[\semM{P[A]} \sdef \{t:\semM{P}|\exi h:\sem{A}\dot \res{t}{o}=h\}\]

%For any implementation or specification object $O$, the program cannot prevent the observation of an operation occurring immediately after its response.
%
%\[\all O, P \dot \all t:\semM{P[O]}\dot \exi t':\semM{P[O]}\dot\\
%\t1 t\sim t' \land \all c:Op\cross Val; i\dot t'_i=obs(c) \implies i > 1 \land t'_{i-1}=res(c) & (\refstepcounter{props}\label{obs}\arabic{props})\]
%%
%where $t \sim t'$  denotes that $t$ and $t'$ are {\em thread equivalent\/}, i.e., when restricted to the events of any one thread they have the same sequence of invocations and responses.

\section{Object refinement}
\label{sec:wmmref}

Correctness of an object is defined from the client program's point of
view. Such a program can only observe changes to {\em program
  variables\/}, i.e., variables that are not defined locally on a
thread or as part of an object.  Let $\obs{t}$ denote the {\em
  observable behaviour\/} of a trace $t$, i.e., the sequence of
observation events of program steps which write to global program
variables. 

A program
$P$ using $C$ on memory model $M$ refines $P$ using
$A$ on $M$ when any observable behaviour of the former is a possible observable behaviour of the latter. We refer to this property as {\em \wmmRef\/}$\!\!$.

\begin{definition}{\WMMRef}
\label{wmmRefdef}

\[P[A] \refsto_M P[C] ~\sdef~ \all t:\semM{P[C]}\dot \exi t':\semM{P[A]}\dot \obs{t'}=\obs{t}\]
\end{definition}

An object implementation $C$ refines an object specification $A$ under weak memory model $M$
if for all possible client programs $P$, $P$ using $C$ weak-memory trace refines $P$ using $A$ under memory model $M$. We refer to this property as \emph{object refinement}.

\begin{definition}{Object Refinement under memory model $M$}
\[ A \refsto_M C ~\sdef~ \forall P \dot P[A] \refsto_M P[C]\]
\end{definition}

If $A \refsto_M C$ we say that $C$ is correct with respect to $A$ under memory model $M$.
%If $A \refsto_M C$ for all $M$, we say $C$ is correct with respect to $A$.

\section{Example application}
\label{sec:ex}

\subsection{Correctness on TSO}
\label{sec:tso}

Consider a spinlock object with operations {\sf acquire}, {\sf release} and {\sf tryAcquire} specified as follows.

%\begin{figure}
\[ \parbox[t]{5cm}{\sf acquire\\ \hspace*{4mm}await(x=1);\\ \hspace*{4mm}x=0 
 } 
 \parbox[t]{4cm}{\sf release\\ \hspace*{4mm}x=1;
 } 
 \parbox[t]{4cm}{\sf tryAcquire\\ \hspace*{4mm}if (x=1) x=0; return 1 \\ \hspace*{4mm}else return 0} 
\]
%\end{figure}
%
\noindent A typical concurrent implementation which is correct on SC is 

%\begin{figure}
\[
 \parbox[t]{6cm}{\sf acquire\\ \hspace*{4mm}while (true) $\{$\\  \hspace*{8mm}if\,(TAS(x,\,1,\,0)=1)\,return\\ 
%\hspace*{8mm}while (x=0) $\{\}$\\ \hspace*{4mm}$\}$
\hspace*{8mm}while (x=0) $\{\}\\
 \hspace*{4mm}\}$
 } 
 \parbox[t]{3.4cm}{\sf release\\ \hspace*{4mm}x=1;
 } 
 \parbox[t]{3.8cm}{\sf tryAcquire\\ \hspace*{4mm}return\,TAS(x,\,1,\,0)}
\]
%\end{figure}
%
\noindent where {\sf TAS(x,a,b)} is the atomic hardware primitive
test-and-set which, when {\sf x} is {\sf a}, sets {\sf x} to {\sf b}
and returns~$1$, and otherwise returns $0$. The {\sf TAS} instruction has a built-in fence to ensure any change it makes to {\sf x} is immediately visible to all threads.

An early version of linearizability on TSO \cite{ifm} proved that this implementation is also correct on TSO. However, using our definition of object refinement we can show that, in fact, it is not correct, and hence that the definition of linearizability in \cite{ifm} is unsound.

\begin{figure}[b]
   \vspace{-8ex}
 \[\qquad \qquad\qquad\parbox[t]{2cm}{~~{\sf T1} \vspace*{2mm}\\ 
      {\sf z =  1;}} 
\parbox[h]{1cm}{\vspace*{2cm}{\resizebox{!}{0.5cm}{$\parallel$}}} 
\parbox[t]{2.5cm}{~~{\sf T2} \vspace*{2mm}\\ 
      {\sf sl.acquire();\\ sl.release();\\ y=z;}} 
%     \parbox[h]{1cm}{\vspace*{2cm}{$\zBig\parallel$}} 
     \parbox[h]{1cm}{\vspace*{2cm}{\resizebox{!}{0.5cm}{$\parallel$}}} 
      \parbox[t]{4cm}{~~{\sf T3}\vspace*{2mm}\\ {\sf await(z=1);\\ w = sl.tryAcquire();}}
 \]
 \caption{Program using spinlock}
  \label{fig:spinlock}
 \end{figure}

Consider the client program in Figure~\ref{fig:spinlock}, which uses a spinlock object {\sf sl}, in which we assume that initially {\sf x = 1} and {\sf z = 0}. One possible trace of this program is \\

\noindent$\lseq inv({\sf T2, sl.acquire), \bot}), res(({\sf T2, sl.acquire,\bot}), obs(({\sf T2, sl.acquire), \bot}), inv(({\sf T2, sl.release),\bot}),$\\ 
$\hspace*{3mm}res({\sf T2, sl.release,\bot}), step({\sf T2, y=0}), step({\sf T1, z=1)}, obs({\sf T1, z=1}), step({\sf T3, await(z=1)}),$ \\
$\hspace*{3mm}inv(({\sf T3, sl.tryAcquire), \bot}), res(({\sf T3, sl.tryAcquire),0}), obs(({\sf T3, sl.tryAcquire),0}), step({\sf T3, w=0}),$\\
$\hspace*{3mm}obs({\sf T3, w=0}), obs(({\sf T2, obs.release),\bot}), obs({\sf T2, y=0})\rseq$\\

\noindent This trace corresponds to thread {\sf T2} acquiring and releasing the lock and reading the initial value of {\sf z}, but not flushing the value written to {\sf x} by the {\sf release} operation until after the other two threads have run to completion. The observable behaviour of the trace is \\

$\lseq  obs({\sf T1, z=1}), obs({\sf T3, w=0)}, obs({\sf T2,y=0})\rseq$\\
~\\
This is not an observable trace of the program running with an object satisfying the specification: if {\sf y=0} then this step and hence {\sf sl.release} on {\sf T2} must have occurred before {\sf z=1} on {\sf T1}, and hence before {\sf sl.tryAcquire} on {\sf T3}. Hence, we do not have object refinement.

The spinlock implementation without the {\sf tryAcquire} operation is, however, known to be correct on TSO
\cite{Sewell:2010:XRU:1785414.1785443}. Again we can show this using our definition. 

The traces of the implementation can be derived from the operational semantics of TSO in \cite{Owens2009}. These show that if an {\sf acquire} has responded (and hence has been observed due to the fence in the {\sf TAS}) then another {\sf acquire} cannot respond until after a {\sf release} on the same core has responded or a {\sf release} on another core has been observed. 
This coincides with what can be observed from the abstract specification. Hence, object refinement holds.

\subsection{Correctness on ARM and Power}
\label{sec:arm}

%On the Power and ARM architectures
%\cite{Alglave:2014:HCM:2633904.2627752,armv8,ColvinFM2018}, writes to
%variables are, like on TSO, local to the core on which they occur and
%are made available to other cores by the hardware or the use of fence
%instructions in the program. However, they are not necessarily made
%available in FIFO order. A write to variable $x$ may be made available
%globally after a write to variable $y$ although the write to $x$
%occurs before the write to $y$ in program order; the write to $x$ and
%the write to variable $y$ seem to occur out of order.

It is easy to show that the spinlock implementation of Section~\ref{sec:tso},
even without the {\sf tryAcquire} operation, 
is not correct on ARM and Power using our definition of object refinement.
For example, consider the client program in Figure~\ref{fig:client-arm} for which we assume 
that initially {\sf x = 1} and {\sf y = 0}.
Following the operational semantics of ARM and Power given in \cite{ColvinFM2018},
one possible trace of this program is\footnote{In the operational semantics of \cite{ColvinFM2018} the placement of observations can be derived from the model of the ``storage subsystem'' which keeps track of which updates to global variables have been seen by which threads.}\\

\begin{figure}[t]
  \vspace{-8ex}
\[
 \qquad \qquad \qquad \quad \qquad\qquad\parbox[t]{2.6cm}{~~{\sf T1} \vspace*{2ex}\\ 
  {\sf sl.acquire();\\ y= y + 1; \\ sl.release()}} 
    \parbox[h]{1cm}{\vspace*{2.6cm}{\resizebox{!}{0.5cm}{$\parallel$}}} 
    \parbox[t]{3cm}{~~{\sf T2}\vspace*{2ex}\\ {\sf sl.acquire();\\ y=y + 1;\\sl.release()}}
\]
\vspace*{-3ex}
\caption{Another program using spinlock}
 \label{fig:client-arm}
\end{figure}

\noindent
$ \lseq 
  inv({\sf (T1, acquire), \bot}), res({\sf (T1, acquire), \bot}), obs({\sf (T1, acquire), \bot}),
   step({\sf T1, y = 1}), $\\
$   inv({\sf (T1, release), \bot}), res({\sf (T1, release), \bot}), obs({\sf (T1, release), \bot}), 
   inv({\sf (T2, acquire), \bot}),$\\
   $res({\sf (T2, acquire), \bot}),
   obs({\sf (T2, acquire), \bot}),
   step({\sf T2, y = 1}), inv({\sf (T2, release), \bot}), res({\sf (T2, release), \bot}), $\\
   $obs({\sf (T2, release), \bot}), obs({\sf T1, y=1}), obs({\sf T2, y = 1})\rseq
$\\
~\\
This trace corresponds to the response of {\sf T1}'s {\sf release} operation being observed before its update to~{\sf y}. This allows {\sf T2}'s {\sf acquire} to occur followed by its update of {\sf y} before {\sf T1}'s new value of {\sf y} is observable by {\sf T2}. Hence, both threads update {\sf y} to 1. 

Since the observable behaviour $\lseq obs({\sf T1, y=1}), obs({\sf
  T2, y=1})\rseq$ of the above trace is not possible using the specification,
the implementation is not correct on ARM or Power: object refinement does not hold. 

 \section{Conclusion}
\label{sec:relwork}

In this paper, we have defined object refinement
which provides a reference point for correctness of concurrent objects
on any weak memory model. This allows notions of
linearizability on weak memory models to be proved sound and
complete.

Observational refinement \cite{Filipovic-LinvsRef2010} and contextual
refinement \cite{Dongol2016} are existing notions of correctness for client/object
systems on SC architectures which coincide with our notion. At the
fundamental level the meaning of a program is found in how it modifies the
observable state.  Where we differ with those frameworks is in how
interactions between the client and object are treated.  In particular, to
handle operations taking effect outside of their invocation/response behaviour,
we introduce the notion of observation events to abstract when a change to
a global variable is visible.  

Doherty and Derrick \cite{DohertyDerrick2016} and Dongol et al.\
\cite{DongolVMCAI2018} address linearizability for weak memory models. Both papers provide arguments that their definitions of linearizability are sound.  At
the level of client/object systems their notions of correctness are similarly
based on changes to observable variables. 

In Doherty and Derrick's framework the semantics of a client operating
with an abstract object specification is not constrained to behaviours
in which the object's specification executes atomically. This does not
affect their results since the client programs are restricted. In
particular, programs like in Figure~\ref{fig:spinlock} which have a
race between the {\sf release} and {\sf tryAcquire} operations are not
allowed. However, our goal is to provide a general notion of
correctness for {\em any\/} client program.

Dongol et al.\ \cite{DongolVMCAI2018} integrate the semantics of
allowed reorderings of invocation and response events with the
reordering semantics framework of Alglave et al.\
\cite{Alglave:2014:HCM:2633904.2627752}. The latter is a well accepted
semantics which can be specialised to TSO as well as ARM and Power.  In their framework, traces of clients using
specifications do not include constraints that the specifications
place on the order of operations, e.g., that successive {\sf acquire}
operations are separated by a {\sf release} operation. These
constraints (captured by the specification order {\sf so}) are added
separately. A consequence of this construction is that correctness
requires linearizability on their notion of traces and an extra
condition (referred to as \textsc{HB-Satisfaction}) to account for the
specification order. In contrast to other work in this area, the
definition of linearizability is \emph{standard} linearizability as
originally defined by Herlihy and Wing \cite{HeWi90}.

Our framework is closer to that original work on linearizability.
Specifically, the allowable order of operations is included in the
traces of the client program calling the specification.  Therefore,
unlike Dongol et al., we do not require an extra condition to handle
specification order.  Using our results from this paper, we would like
to investigate whether standard linearizability alone
can be proven sound and complete for any client program and weak memory model.

\vspace*{2ex}
\noindent{\bf Acknowledgement} \sloppy This work was supported by Australian Research Council Discovery Grant DP160102457.

\bibliographystyle{eptcs}
\bibliography{references}

\end{document}